\documentclass[10pt]{article}

\usepackage[utf8]{inputenc}		
\usepackage[affil-it]{authblk}

\usepackage[T1]{fontenc}
\usepackage{url,color}
\usepackage{graphics,amsfonts}
\usepackage{epstopdf}
\usepackage[pdftex]{graphicx}
\usepackage[cmex10]{amsmath}
\usepackage{braket}
\usepackage[explicit]{titlesec}
\usepackage{hyperref}
\usepackage[nameinlink]{cleveref}
\usepackage{enumitem}
\usepackage{blindtext}
\usepackage{siunitx}
\usepackage{float}
\usepackage[font=small,labelfont=bf]{caption}
\usepackage{subcaption}
\usepackage{abstract}
\usepackage[top=1.5cm, bottom=2cm, right=1.6cm,left=1.6cm]{geometry}

\usepackage[
backend=biber,
style=nature,
sorting=none,
citestyle=numeric-comp, 
doi = false,
url = false,
maxnames = 6
]{biblatex}

\makeatletter
\def\@maketitle{%
	\newpage
	\null
	\vskip 2em%
	\begin{center}%
		\let \footnote \thanks
		{\LARGE \textbf \@title \par}%
		\vskip 1em%
		{\small
			\lineskip -1.5em%
			\begin{tabular}[t]{c}%
				\@author
			\end{tabular}\par}%
		{\small (Dated: \@date)}%
	\end{center}%
	\par
	\vskip .5em}
\makeatother

\titleformat{\section}
{\normalfont\bf\center}{\thesection}{1em}{\MakeUppercase{#1}}
\titleformat{\subsection}
{\normalfont\bf\small\center}{\thesection}{1em}{#1}
\titleformat{\abstract}
{\normalfont\bf\center}{\thesection}{1em}{{#1}}

\title{{Trap-dependent current suppression of optically excited III-V nanowires at cryogenic temperatures}}

\author[1]{Myriam Rihani\thanks{Corresponding author: mrihani@phys.ethz.ch}}

\author[2,1]{Cristina Martinez-Oliver}

\author[1]{Markus A. Scherrer}

\author[1]{Heinz Schmid}

\author[3,4]{Kirsten E. Moselund}

\author[3,4]{Simone Iadanza}

\affil[1]{IBM Research, Säumerstrasse 4, 8803 Rüschlikon, Switzerland}
\affil[2]{University of Glasgow, James Watt School of Engineering, Glasgow G12 8QQ, United Kingdom}
\affil[3]{Paul Scherrer Institut, Forschungsstrasse 111, 5232 Villigen, Switzerland}
\affil[4]{École Polytechnique Fédérale de Lausanne (EPFL), Route Cantonale, 1015 Lausanne, Switzerland}

\date{\today}

\begin{document}
	\maketitle
	
	\begin{abstract}
		The advancement of quantum technology networks necessitates high-speed, low-thermal load, and minimal-noise communication links between cryogenic and room-temperature components. At the heart of modern telecommunication, lay optical interconnects allowing for large data transfer capabilities via optical fibers. However, cryogenic photonic technologies remain largely unexplored and require a detailed understanding of material behavior and defect dynamics at low temperatures. In this work, we present the first comprehensive study of integrated III–V heterostructures operating at cryogenic temperatures down to 5 K. Using an integrated n-InP/i-InGaAs/p-InP/p-InGaAs stack monolithically grown on silicon, we identify a temperature-dependent current-lowering mechanism arising from trap states becoming increasingly active below 140K. We demonstrate for the first time that these traps can be equivalently excited and controlled through either thermal or optical energy, revealing a dual modulation mechanism. These findings provide new insights into carrier transport and defect behavior in III–V heterostructures at cryogenic temperatures, advancing the field of cryogenic photonics and offering a non-destructive approach for identifying and characterizing material impurities in integrated quantum and optoelectronic devices.
	\end{abstract}
	
    \newpage
\section{Introduction}

The exploration of microsystems at cryogenic temperatures have experienced significant growth in the past decade, driven largely by rapid advances in quantum technologies and quantum computing\cite{van_deventer_towards_2022, charbon_cryogenic_2021}. Most quantum processors based on superconducting or spin quantum bits (qubits) require cryogenic temperatures to ensure device functionality, environmental stability and suppress noise. Protocols such as quantum error correction \cite{bravyi_high-threshold_2024, hetenyi_creating_2024, fowler_surface_2012} require reliable, real-time, and parallel control of large, interconnected qubit lattices at millikelvin (mK) temperatures. As these systems scale up, integrating increasing numbers of qubits on a single chip \cite{krinner_engineering_2019, gao_establishing_2025, acharya_quantum_2025, kim_evidence_2023}, their manipulation and measurement via control and readout electronics face growing bottlenecks, including system footprint, thermal load, I/O limitations, and latency \cite{reilly_challenges_2019}.

There have been efforts to tackle these problems. Recent research focused on the co-integration of qubits with their electronics, utilizing the foundry-level know-how of complementary metal–oxide–semiconductor (CMOS) technologies \cite{bartee_spin-qubit_2025}. 
However, the stringent power consumption constraints in dilution refrigerators at the qubit stage is a critical limitation \cite{pauka_cryogenic_2021}. Multiplexing strategies can help in reducing the number of RF cables across the dilution fridges, however these schemes are limited by cross-talk and bandwidth of electronic cabling \cite{chen_multiplexed_2012, heinsoo_rapid_2018}.

To address these constraints, the replacement of coaxial cables for data transfers with photonic links utilizing silica optical fibers has been proposed \cite{awschalom_development_2021}. Optical fibers have ultra-low signal losses (0.2 $\textrm{dB}\cdot\textrm{km}^{-1}$), compared to coaxial cables at gigahertz frequencies (3.0 $\textrm{dB}\cdot\textrm{m}^{-1}$) and exhibit extremely low passive heat loads (pW range), orders of magnitude lower than RF coaxial cables \cite{krinner_engineering_2019,youssefi_cryogenic_2021}. Optics and photonic technologies offer advantages not only in low propagation loss, hence lower cross talk, but also with high speeds (beyond Tbps) and large bandwidths. Consequently, new schemes using photonic links for quantum computing are currently being proposed in the literature, in which optical interconnects are operated at cryogenic stages \cite{youssefi_cryogenic_2021, arnold_all-optical_2025, joshi_scaling_2022}. 

Photodetectors play a fundamental role in electro-optic hybrid architectures, serving as converters of optical signals to electrical ones for transmission through optical fibers. A first demonstration of superconducting qubit control and readout via a photonic link was achieved using a commercial InGaAs photodiode at the mK stage of a dilution refrigerator \cite{lecocq_control_2021}. The monolithic integration of such photonic devices with CMOS technologies represents a critical step toward realizing the full benefits of photonic links for cryogenic quantum systems. Silicon is not only the foundation for modern CMOS electronics, but also increasingly exploited as the waveguiding medium in silicon photonics for the near-infrared C and O band data-communication wavelengths. This requires the introduction of alternative materials to serve as integrated photodetectors (Ge and III-V compounds) and light sources (III-Vs). Integration of III-V compounds into silicon remains difficult due to significant lattice mismatch between the two materials. As a result, the development of reliable fabrication processes for monolithic integration of III–V materials on silicon is an active area of research\cite{borg_vertical_2014, schmid_template-assisted_2015, borg_facet-selective_2018, nanwani_monolithic_2025}. Using an epitaxial growth method called template-assisted selective epitaxy (TASE), a first monolithic integration of III-V nanowires on Si has been demonstrated and operated in the GHz regime \cite{wen_waveguide_2022}. A crucial step for quantum schemes remains the investigation of these photodetectors at cryogenic temperatures. 

Here, we present a first investigation of the performance of a waveguide-integrated III-V nanowire from room to cryogenic temperatures, down to 5K. The III-V heterostructures are p-i-n junctions consisting of a stack of n-InP/i-InGaAs/p-InP/p-InGaAs monolithically grown on a silicon platform using TASE. Their very small footprint - active area of $~\lambda^3/10^3$, potentially unlocking sub-fF capacitance ranges, offer a great advantage for low-power and high-speed applications \cite{mauthe_high-speed_2020}. The electro-optic properties of these nanowires were studied for varying temperatures from 300K to 5K by measuring their dark current and photocurrent characteristics, proving their reliable operation under strong temperature changes. This work resulted in the observation and study of thermally and optically dependent trap behaviors near band-gap edges. Moreover, in this work we demonstrate for the first time trap-dependent dynamics leading to optically induced photodiode current suppression phenomena below dark current ranges in the cryogenic regime (i.e. a reduction of photocurrent with increasing optical excitation energy, up to a certain optical threshold). The mechanism investigated here, opposite to conventional photodiode operation, may be employed in future accurate optical cooling applications and non-destructive photodetector defect concentration diagnostics.

\begin{figure}[h!]
    \centering
    \includegraphics[width=0.9\linewidth]{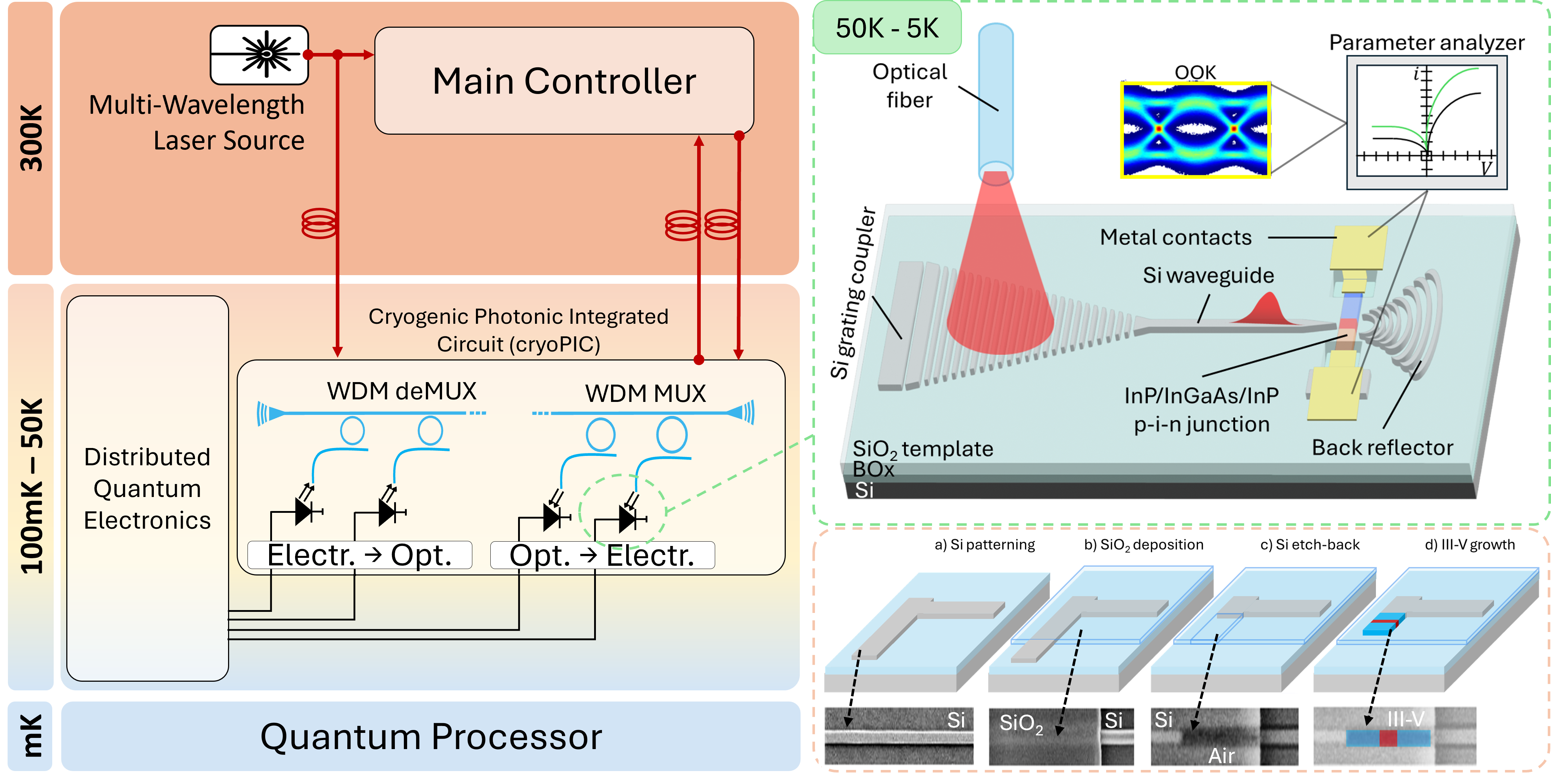}
    \caption{Proposed hybrid scheme using cryogenic photonic integrated circuits (cryoPIC) for quantum computing. Information is communicated between the distributed control electronics and the room temperature controller via optical fibers. Modulated light is coupled to the cryoPIC using grating couplers and directed to the electo-optic device via integrated silicon ridge waveguides.}
    \label{fig:intro}
\end{figure}

    \section{Results and Discussion}

\subsection{III-V heterostructures}
\begin{figure}[h!]
    \centering
    \includegraphics[width=1\linewidth]{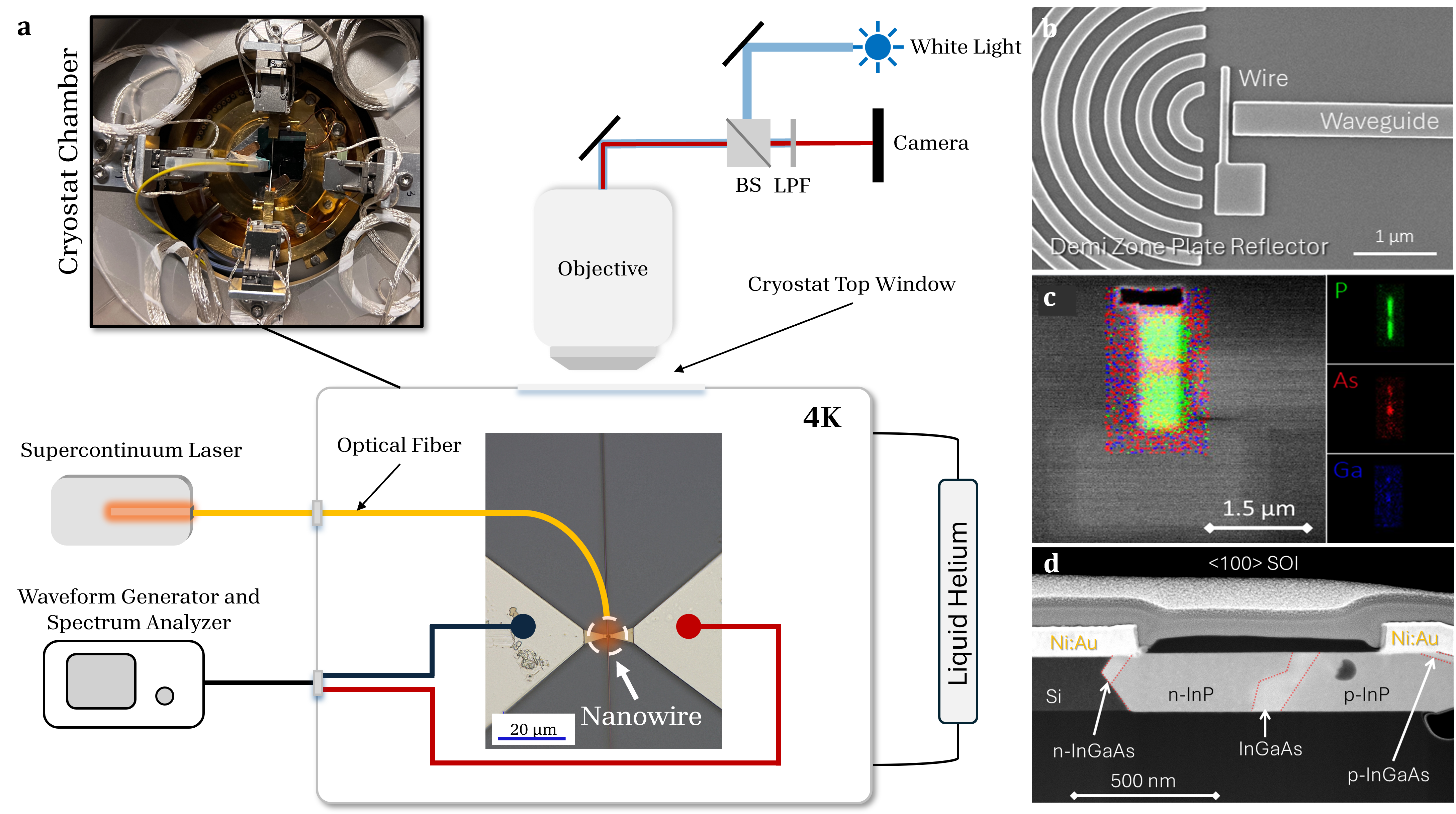}
    \caption{(a) Experimental set-up for cryogenic characterization and study of devices. Liquid helium cooled cryostat (chamber picture in the inset) enables cooling down to 4K. (b) SEM of fabricated waveguide-coupled III-V nanowire with integrated half-zone plate reflector to optimize optical absorption. (c) EDX image of device, highlighting local distribution of P, As and Ga compositions. (d) TEM images of the fabricated heterostructures for $<100>$ SOI orientation. }
    \label{fig:eo_temp}
\end{figure}

The devices investigated in this work are III–V heterostructures monolithically integrated onto a silicon-on-insulator (SOI) platform, enabling efficient coupling to silicon optical waveguides via grating couplers (Fig.\ref{fig:intro}). The heterostructures form p–i–n junctions composed in a stack of n-InP/i-InGaAs/p-InP/p-InGaAs. Due to the strong lattice mismatch between III-V material and silicon, we use template-assisted selective epitaxy combined with in-situ doping to grow the nanowire stack in the silicon platform. Further details on device fabrication can be found in \cite{schmid_template-assisted_2015,mauthe_high-speed_2020}. The structure is patterned onto the SOI wafer using electron-beam lithography and etched with inductive-coupled plasma with HBr chemistry (Fig \ref{fig:intro}). The structure is then uniformly encapsulated with a $\rm{SiO_2}$ layer before crating an opening at the device region to selectively remove the silicon with wet-etching while leaving behind a thin silicon seed layer, promoting III-V nucleation. The III–V active layers are then grown by metal-organic chemical vapor deposition and the Ni:Au contacts are patterned and deposited on the heterostructures using EBL followed by a metal lift-off process. Due to InGaAs absorption properties, the nanowires exhibit high spectral response from 1200nm to 1700nm. For lattice-matched $\mathrm{In_{0.53}Ga_{0.47}As}$, the absorption coefficient is maximized in the 1300-1400nm range. SEM images as well as TEM inspection are both shown in Fig.\ref{fig:eo_temp}b,c,d.

\subsection{Temperature dependent electro-optic characterization of III-V heterostructures}
\label{subsec:temp_eo}

The electro-optic properties of the on-chip photodetectors can be characterized by applying a bias across the device and measuring the current-voltage (IV) characteristics with and without optical excitation using a waveform generator and spectrum analyzer. To perform the temperature study, the chip is placed inside a liquid-helium cryostat with temperature control ranging from 5K to 300K. The devices are optically excited using a supercontinuum laser (640-2000nm, 2W total power - spectral information in experimental methods Sec. \ref{sec:laser_info}) and biased electrically using a waveform generator and spectrum analyzer (Fig.\ref{fig:eo_temp}a). Although the photodetectors are coupled to a waveguide, the associated grating couplers have a relatively narrow transmission. Therefore, for the spectral studies we choose to measure the devices under direct illumination via an optical fiber as illustrated in Fig.\ref{fig:eo_temp}a. A further discussion on the measurement set-up is done in the experimental methods \ref{sec:exp}. 

\begin{figure}[h!]
    \centering
    \includegraphics[width=0.8\linewidth]{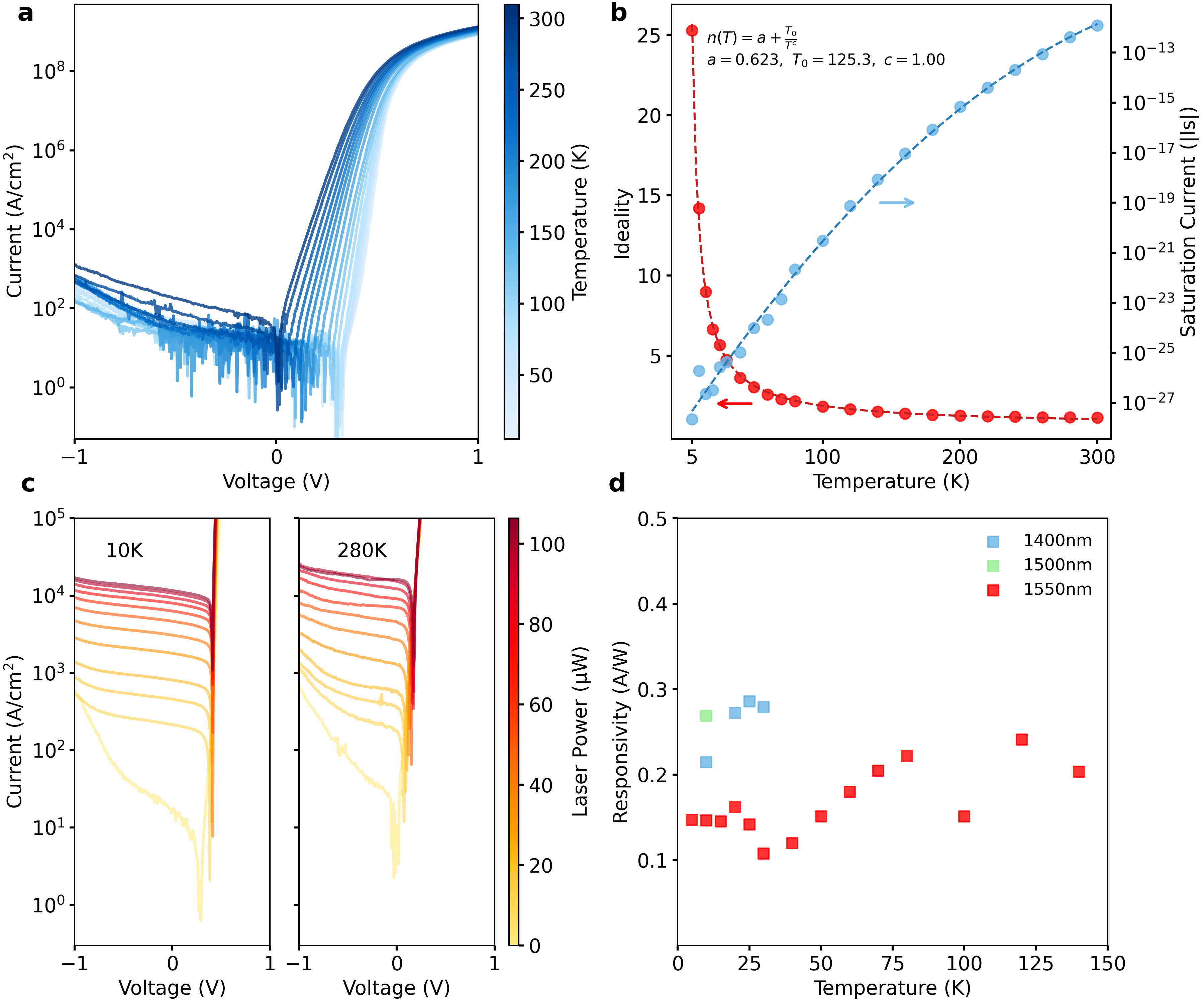}
    \caption{(a) Dark-current IV curves at temperatures between 300 to 5K. (b) Ideality (red) and saturation current (blue) data extracted from dark-current measurement vs temperature. The data points were fitted using the Levine model for the ideality factor and a quadratic fit for the saturation current. (c) Photocurrent IV characterization at 10K (left) and 280K (right) at varying laser powers at 1550nm wavelength. All the IV curves are plotted in absolute value of the current. (d) Responsivity of nanowire vs Temperature for different wavelengths.}
    \label{fig:dark_vs_temp}
\end{figure}

The fabricated \textit{p-i-n} heterostructures exhibit diode-like IV curve behaviors, with two different slopes for reverse and forward biasing of the junction. With no optical illumination, the dark current flowing through the device is dependent on thermal excitations and environmental noise. At lower temperatures, thermal noise becomes less important, resulting in a decrease of the total dark current level (Fig.\ref{fig:dark_vs_temp}a). 
Figures of merit of photodiodes are their ideality factor and saturation currents which can be extracted from the measured data using the forward bias slope and its interception with the y-axis (Fig.\ref{fig:dark_vs_temp}b) \cite{schroder_carrier_2005}. 

The extracted ideality factor exhibits a strong temperature dependence, increasing significantly as the temperature is reduced. Such behavior has been widely reported in Schottky contacts and is commonly attributed to lateral inhomogeneities in the Schottky barrier height at the metal–semiconductor interface, as well as the presence of interface states and defects whose contribution becomes more pronounced at low temperatures \cite{mamor_interface_2009, ahaitouf_interface_2012}. As a result, ideality factor values substantially exceeding the typical range of 1–2 observed at room temperature can be obtained. To quantitatively describe the temperature dependence of the ideality factor, the data is fitted using the Levine model, expressed as $n=a+T_0/T^c$ ($T_0$ the pseudo-temperature and $a$ and $c$ are temperature-insensitive parameter) \cite{levine_schottky-barrier_1971}. This model has been widely employed as a phenomenological framework to capture non-ideal transport behavior in Schottky junctions.

It is worth noting that the temperature range explored in this work extends well below that typically reported for comparable devices, where measurements are often limited to temperatures above 30–50K \cite{akkal_electrical_2004, ahaitouf_interface_2012, mamor_interface_2009, dalapati_analysis_2020}. Access to this cryogenic regime reveals transport characteristics that are not observable at higher temperatures. The Levine model is derived under the assumption of thermally activated carrier emission and a current dominated by recombination via defect states. At very low temperatures, thermal emission from traps is progressively suppressed, and charge transport is expected to transition toward tunneling-assisted mechanisms. These processes can still produce a quasi-exponential $I(V)$ characteristic, resulting in very large apparent ideality factors that no longer reflect a single recombination mechanism \cite{dalapati_analysis_2020}. The observed increase in ideality factor at low temperatures is therefore consistent with the non-ideal transport behavior inferred from the Richardson plot discussed in Section \ref{sec:temp_var} with Fig.\ref{fig:fig4}b.

The optical behavior of the detector at different temperatures, 10K and 280K (Fig.\ref{fig:dark_vs_temp}c), shows a strong photocurrent response when illuminating with increasing laser power. The responsivity $(\mathrm{R} = \frac{\mathrm{dI_{ON/OFF}}}{\mathrm{dP_{optical}}})$ of the photodiodes at different temperatures can be extracted using the generated photocurrent at 20$\mu \mathrm{W}$ and 100$\mu \mathrm{W}$ laser power at a diode bias of $-0.8\mathrm{V}$, away from the noise and anomalies at lower bias and photocurrents. At higher wavelengths, closer to the InGaAs absorption window, the responsivity is higher, up to 28\% for 1400nm compared to the 15\% at 1550nm (Fig.\ref{fig:dark_vs_temp}d). 

\subsection{Observation of temperature-dependent current lowering} \label{sec:temp_var}
\begin{figure}[h!]
    \centering
    \includegraphics[width=0.8\linewidth]{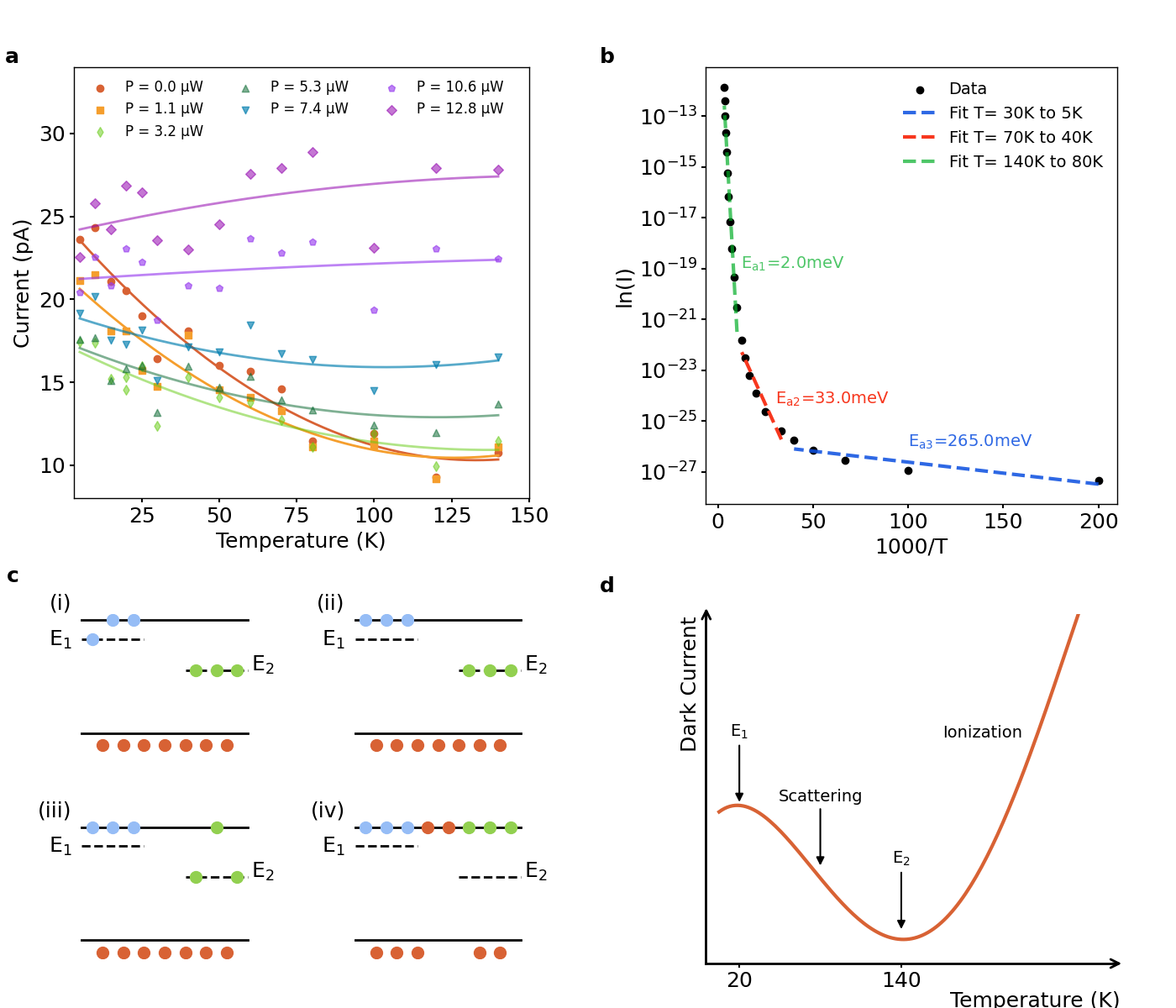}
    \caption{(a) Current vs Temperature plot for varying low optical excitation highlighting the current lowering up to 140K. Measurements were done at 1550nm with varying optical powers. (b) Richardson plot extrapolated from experimental data with corresponding activation energies at different temperature regimes. (c)-(d) Schematic explaining the fabricated III-V heterostructures behavior at different temperature regimes, highlighting the presence of shallow ($\mathrm{E_1}$) and deep ($\mathrm{E_2}$) defects that progressively get excited by thermal noise.}
    \label{fig:fig4}
\end{figure}

We reveal interesting deviations from conventional physical behavior of the photocurrent at temperatures lower than 140K. At a fixed bias of $-1\mathrm{V}$, the dark current decreases as the temperature increases (red curve in Fig.\ref{fig:fig4}a). A similar behavior, measured in resistivity has first been observed for bulk InGaAs, which has been associated to traps occupying specific energy levels within the band-gap of the material \cite{tian_thermal_2019}. At very low temperatures (around 5K), the device dark current is due to conduction and valence band related defects, and an initial slight increase in dark current is observed (Fig \ref{fig:fig4}d). Increasing temperature further increases the thermal energy of the system until it is enough to excite shallow traps from the energy levels in the band gap ($E_1$) closer to the conduction band, which in turn starts to reduce the dark current. This keeps occurring, partly balanced by carrier scattering further activating the traps in the deeper levels of the bandgap (still reducing the dark current), until no trap is left for excitation to the conduction band (Fig.\ref{fig:fig4}c(i)-(ii)-(iii)). 

After trap excitation saturation, increasing temperature further leads valence carriers excitation to the conduction band, i.e. the conventional behavior for dark current (ionization in Fig.\ref{fig:fig4}d \& Fig.\ref{fig:fig4}c(iv)). Using the fit of the ideality factor and saturation current extracted in Sec. \ref{subsec:temp_eo}, the Richardson plot of the extrapolated dark current with corresponding activation energies at different temperature regimes is shown in Fig.\ref{fig:fig4}b. The activation energies of the shallow and deep traps (2meV and 265meV respectively) are comparable to the bulk InGaAs analysis \cite{tian_thermal_2019}, supporting the measurements in this work. This further entails the presence of defects and therefore tunneling-assisted current that are not accounted for in the Levine model, which explains the exponential increase in ideality factor with a decrease of temperature.  

This current-lowering effect can be generated not only by temperature but also via optical excitation of the nanowire photodetectors. The different curves in Fig.\ref{fig:fig4}a represent the photodetector current in the dark regime (P=0 $\mu W)$) and  for increasing laser powers at an operating wavelength of 1550nm. The dark current follows the trend of Fig.\ref{fig:fig4}d and \cite{tian_thermal_2019} when applying no laser power. By increasing laser power at a constant temperature the device photocurrent decreases below the dark current levels (this effect becomes less pronounced with increasing temperature until it disappears at 140K). The measured photocurrent decrease follows a parabolic trend with temperature, forming a dip that becomes more pronounced from $0\mu W$ to $1.1\mu W$ laser power and progressively flattens out at higher optical powers, until disappearing at $12.8\mu W$ laser power (when the photocurrent only increases, as conventionally shown in literature). This current lowering effect leads to the crossing between the dark current curve (red fitted line in Fig.\ref{fig:fig4}a) with the other measured photocurrent curves. This crossing occurs at progressively decreasing temperatures for increasing optical power, hinting towards an energy equilibrium mechanism between thermal and optical energy. 

\subsection{Investigation of optically-dependent energy threshold}

\begin{figure}[!ht]
    \centering
    \includegraphics[width=0.8\linewidth]{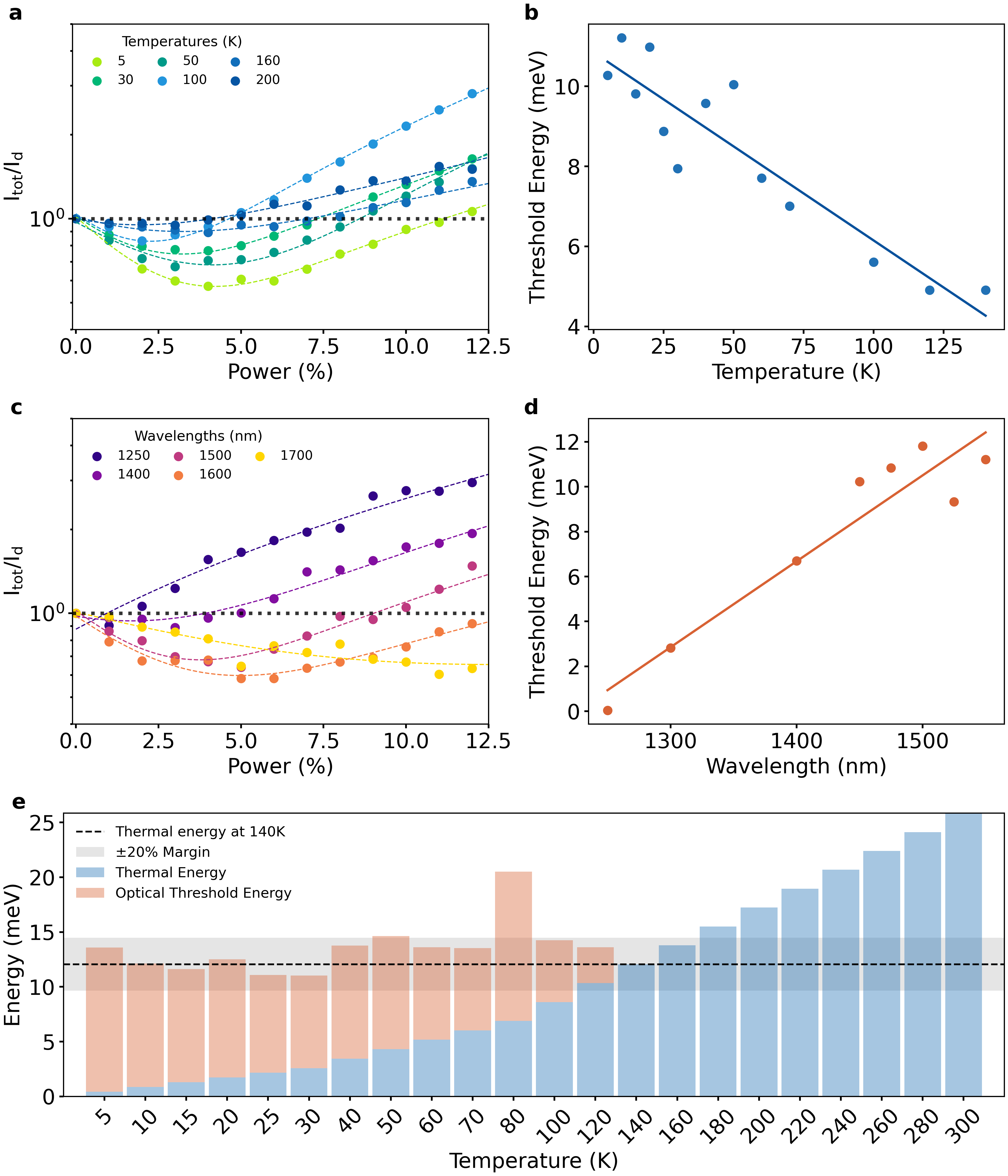}
    \caption{(a) Current vs laser power percentage at 1550nm for different temperatures. The current for each temperature is normalized over the dark current and the dotted black line represents unity. (b) Threshold energy (corresponding to optical energy of the current lowering minima) vs temperature. (c) Current vs laser power percentage at 30K for different wavelengths. The current at each wavelength is normalized by the dark current at 30K. The unity line is represented by the black dashed line. (d) Threshold energy vs Wavelength at 30K with linear fit of the data. (e) Bar plot of Energy vs Temperature. Thermal energy (blue bars) and optical threshold energy (red bars) are summed showing a total energy equivalent to thermal energy at 140K (black dashed line).}
    \label{fig:fig5}
\end{figure}

\begin{figure}[h!]
    \centering
    \includegraphics[width=0.5\linewidth]{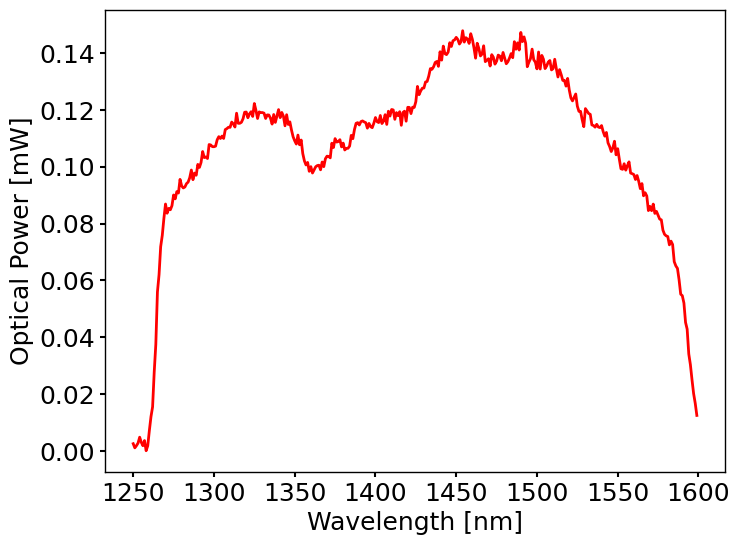}
    \caption{Optical power spectrum at 100\% laser output, accounted for in the optically-dependent energy threshold analysis. The variation of laser power is taken in percentage of the total output laser power in all sections.}
    \label{fig:laser}
\end{figure}

The current lowering dependence with optical energy at different temperatures is shown in Fig.\ref{fig:fig5}a. The photocurrent, normalized to the dark current for each temperature, decreases with increasing applied optical power (here at 1550nm wavelength) until reaching a minimum. By increasing the temperature, the current minimum shifts to lower optical powers. This current minimum point at each temperature occurs at the optical threshold energy, which is the energy required for the current lowering effect to disappear at temperatures below 140K. The threshold energy against temperature is plotted in Fig.\ref{fig:fig5}b showing a clear linear trend. Similarly at a fixed temperature (30K), the current dependence on optical power is plotted at different wavelengths, hence different effective optical energies, (Fig.\ref{fig:fig5}c). At lower wavelengths (higher photon energies), photocurrent increases with increasing optical power input percentage of the laser (the maximum 100\% power in mW of the laser for different wavelengths is given in Fig.\ref{fig:laser} in section \ref{sec:exp}). Lowering photon energy (increasing wavelength) leads to this anomalous current minimum that progressively shifts at higher laser powers, until at 1700nm there is no longer any photocurrent increase ($\mathrm{In_{0.53}Ga_{0.47}As}$ absorption limit). Fig.\ref{fig:fig5}d. shows the measured optical energy required for ${I_{tot}}/{I_{d}}$ to cross the normalization point at the end of the photocurrent dip in Fig.\ref{fig:fig5}c against wavelength at a constant temperature of 30K (constant thermal energy). The measured data exhibits a linear trend, as expected from a photon rate increase proportional to $\lambda$ associated to the increasingly higher laser powers (required at higher wavelengths to reach the same ${I_{tot}}/{I_{d}}$ baseline after the photocurrent dip).
This mechanism relating the thermal and optical energy to the current lowering effect observed in this experiment can be better explained by plotting the total energy vs temperature, as shown in Fig.\ref{fig:fig5}e. Here, the thermal energy corresponding to the operating temperature is added to the optical energy (orange box) associated to the increasing laser power required to increase the photocurrent to the dark current level after the photocurrent drop below thermal noise shown in Fig.\ref{fig:fig5}a. At each temperature, the sum of optical and thermal energy equals the thermal energy at 140K, temperature at which the system has already enough heat for the traps to be completely excited to conduction band and their associated the current lowering mechanism stops being observed.

    \newpage
\section{Conclusion}

In this work, to the best of our knowledge  we perform for the first time a complete study of a chip-integrated III-V heterostructures at cryogenic temperatures, down to 5K. We have demonstrated for the first time a heat-dependent and optical energy-dependent current lowering mechanism in a n-InP/i-InGaAs/p-InP/p-InGaAs stack, monolithically grown in silicon. This phenomenon suggests that the presence of defects and impurities becomes dominant on the photodiode performances at low temperatures (below 140K) and small optical powers (a few 10µW). The excitation of these defects can be manipulated bilaterally with specific combinations of temperature and optical injection, enabling the possibility to tune the integrated photodiode characteristics and switch between high photocurrent generation regimes and photo-induced below dark current regimes. This work advances knowledge on photodetection in cryogenic environments, fundamental for qubit readout applications in quantum technology and provide a novel  mechanism to be employed for photodetector “optical cooling” below dark current values or as a non-destructive method for defect concentration monitoring  in such devices in post-fabrication.
    \section{Experimental Methods} \label{sec:exp}
\subsection{Measurement Set-Up}
The electro-optic properties of the on-chip photodetectors can be characterized by applying a bias across the device and measuring the current-voltage characteristics with and without optical excitation from a fiber positioned on top of the device. To perform the temperature study, the photodetectors are placed on a sample holder stage inside a vacuum chamber equipped with x-y-piezo for precise alignment. The sample can then be cooled to temperatures between 5K-300K using liquid Helium and heated up with a CryoVac TIC 500 temperature controller. The devices were optically excited using a high-power supercontinuum laser (640nm-2000nm). 

\subsection{Laser power Spectrum} \label{sec:laser_info}

\subsection{Ideality factor and saturation current}\label{sec:exp_ideality}
The ideality factor and the saturation current can be extracted from the measured data using the forward bias slope \cite{schroder_carrier_2005}. The dark current equation can be expressed as : 
\begin{equation}
    I_d = I_s\exp{(\frac{qV}{nk_BT})} \cdot(1-\exp{(\frac{-qV}{k_BT}}))
\end{equation}
The quantities $k_B$, $T$, and $q$ represent the Boltzmann constant, temperature and electron charge. The semi-log of the IV curve would then be linear for values of V where $V >> k_BT/q$. Hence, the current equation at this region can be expressed as, in semi-log scale:
\begin{equation}
\label{eq:dark_current_slope}
    \log (I_s\exp{(\frac{qV}{nk_BT})} ) = \log (I_s) + \frac{q}{nk_BT} \cdot V
\end{equation}
From there, one can extract the ideality and the saturation current using the slope and the intercept to the y-axis of equation \ref{eq:dark_current_slope}. The slope of each IV curves at different temperatures between $10^{-10}$pA and $10^{-8}$pA is used in the measured data and extrapolated to find the intersection with the y-axis \cite{schroder_carrier_2005}. The saturation current is fitted by a quadratic fit. 

The ideality factor was extracted from the forward-bias $I(V)$ characteristics measured over a wide temperature range. A temperature dependence of the extracted ideality factor was observed, particularly at low temperatures. To parameterize this behavior, the temperature dependence of the ideality factor was fitted using the Levine model, expressed as

\begin{equation}
\label{eq:id_model}
    n = a + T_0/T^{c}
\end{equation}

where $T_0$ is a pseudo-temperature, and $a$ and $c$ are temperature-independent fitting parameters \cite{dalapati_analysis_2020}

Although the Levine model is formally derived under thermally activated transport conditions, it was applied here as a fitting framework to quantify the strong temperature dependence of the experimentally extracted ideality factor \cite{levine_schottky-barrier_1971, mamor_interface_2009, ahaitouf_interface_2012}.

	\section*{Acknowledgements}
    The authors thank the Swiss National Science Foundation for the NCCR SPIN INSPIRE Potentials Fellowship. The authors also thank the European Union’s Horizon 2020 research and innovation program. The authors acknowledge the Cleanroom Operations Team of the Binning and Rohrer Nanotechnology Center for their help and support.

	\section*{Contributions}
    M.R. and S.I. conducted the experiment and study. C.-O.M. fabricated the sample and contributed to the measurements and discussion. M.A.S. and H.S. helped develop the experimental set-up. K.E.M., M.A.S. and H.S. contributed to result discussion. M.R. analyzed the data, with input from S.I. and wrote the original draft of the manuscript. K.E.M. and S.I. supervised the project.
	
	\section*{Competing interest}	
	The authors declare no competing interests.
	
	\printbibliography
	
\end{document}